# Collision-induced Hopf-type bifurcation reversible transitions in a dual-wavelength femtosecond fiber laser

RUNMIN LIU,[1] DEFENG ZOU,[1] SHUANG NIU,[1] YOUJIAN SONG,[1,2] AND MINGLIE HU[1,3]

[1]*Ultrafast Laser Laboratory, Key Laboratory of Opto-electronic Information Science and Technology of Ministry of Education, School of Precision Instruments and Opto-electronics Engineering, Tianjin University, 300072 Tianjin, China*
[2]*yjsong@tju.edu.cn*
[3]*huminglie@tju.edu.cn*

**Abstract:** Collision refers to a striking nonlinear interaction in dissipative systems, revealing the particle-like properties of solitons. In dual-wavelength mode-locked fiber lasers (MLFLs), collisions are inherent and periodic. However, how collisions influence the dynamical transitions in the dual-wavelength mode-locked state has still not been explored. In our research, dispersion management triggers the complex interactions between solitons in the cavity. We reveal the smooth or reversible Hopf-type bifurcation transitions of dual-color soliton molecules (SMs) during collision by real-time spectral measurement technique of time-stretch Fourier transform (TS-DFT). The reversible transitions from stationary SM to vibrating SM, revealing that cavity parameters pass through a bifurcation point in the collision process without active external intervention. The numerical results confirm the universality of collision-induced bifurcation behavior. These findings provide new insights into collision dynamics in dual-wavelength ultrafast fiber lasers. Furthermore, the study of intermolecular collisions is of great significance for other branches of nonlinear science.

## 1. Introduction

As one of the most fascinating nonlinear interactions, soliton collisions have been extensively investigated in the fields of Bose-Einstein condensates (BECs) [1,2], cell biology [3], fluid mechanics [4], nonlinear fiber optics [5-10], etc. In an integrable Hamiltonian system governed by the nonlinear Schrödinger equation (NLSE), the soliton maintains its original envelope stability after collision [1,11]. However, the loss of integrability results in the annihilation or fusion of soliton pairs [12]. Another research on the interaction between atomic BECs and light field in an optical ring cavity demonstrates that the outcomes of collisions depend on the approaching velocity of solitons, where medium velocity collision induces a severe scattering and partial annihilation of BEC atoms (the typical behavior for dissipative solitons) [13]. That is, soliton collisions exist in a wide range of non-integrable and non-conservative systems. In a dissipative system, the collision dynamics are more complicated because of another set of nonlinear gain and loss balance. For example, non-chemotactic mutants of the cellular slime mould exhibiting soliton-like structures pass through one another in multicellular movement [3].

Moreover, due to the typical dissipative properties of MLFLs, numerous efforts have been devoted to research their internal complex nonlinear dynamics [14-18]. However, limited by the compromise of scanning speed and accuracy, conventional measurement methods can only exhibit time-averaged results, which are insufficient to reveal the complex physical mechanism in the oscillator. With the rapid development of real-time spectroscopy technology, the perfect match between TS-DFT and fiber lasers provides an optimal solution to explore the transient processes of soliton dynamics. Inspired by the pioneering observation of the internal motions of SMs, researchers are keen to demystify the nonlinear interactions between solitons, such as the buildup of solitons [19-20], stationary [22-24] and vibrating SMs [25-27], soliton explosions [28-32], rogue wave [33-35], etc.

In MLFLs, collisions originated from differences in phase and group velocity between solitons. Generally, collisions are triggered randomly and instantaneous responses to the change in cavity parameters. The peak power clamping effect suppresses the excessive accumulation of energy, resulting in the generation of multi-pulses or self-organized structures of SMs [36]. Besides, a soliton pair and a single soliton (SS) appear simultaneously with certain cavity parameters and must collide at some stage due to their group velocity difference [7]. The different group velocities of the collision pulses may be caused by stimulated Raman scattering that is shifting differently their carrier frequencies [6]. With the influence of gain dynamics, the fast gain response also enables different soliton organized structures to exhibit different drift velocities [9]. Another experimental result identifies that the increase of intracavity energy drives the stable double pulses to generate the relative velocity difference, and then the collision triggers the extreme event of the soliton explosion [37].

However, the collisions mentioned above, which are observed in single-wavelength MLFLs, appeared randomly. In recent years, due to the application of dual-comb sources in optical fiber sensing, absolute distance measurement, molecular spectroscopy [38-40], etc., dual-wavelength MLFLs have gradually come into views of researchers. In fundamental scientific researches, the inherent periodic collisions are a perfect entry point for the researches of complex nonlinear interactions in dual-wavelength MLFLs. With the support of TS-DFT technique, various nonlinear phenomena during collision have been revealed. In the buildup process of dual-wavelength mode-locking, soliton collisions facilitate noise to evolve into a mode-locked state at the new wavelength [41]. In an anomalous-dispersion dual-color fiber laser, the collisions result in dispersive wave shedding associated with Kelly sidebands and reorganization of multi-pulse SMs [42,43]. In addition, periodic collisions also cause periodic soliton explosions and the generation of dissipative rogue waves [44]. Despite the buildup process and collision-induced exotic dynamics of dual-wavelength MLFLs have been investigated, there is still an issue worth considering: How bifurcation behaviors ubiquitous in dissipative systems are influenced by collisions? It is well known that the bifurcation transition is reversible, such as the transition between fixed-point attractors and limit-cycle attractors [11]. However, the bifurcation reversible transitions during collision have not yet been discovered. Exploring the influence of collisions on Hopf-type bifurcations is of great significance for investigating the complex nonlinear dynamics in dual-wavelength MLFLs.

Here, we reveal various nonlinear phenomena during collision in a dual-wavelength dispersion-managed mode-locked fiber laser based on nonlinear polarization rotation (NPR) technique. The whole and clear collision process has been observed by DFT real-time spectral measurements and extra-cavity delay line. Firstly, we observe the details of collision between SSs. Subsequently, the smooth transitions of dual-color SMs are observed with the higher pump power and appropriate PC orientations. Remarkably, no matter the stationary SM or the vibrating SM, the smooth transition is reflected in that the collision does not change the mode-locked state. With the same PC orientations and pump power, reversible bifurcation transitions during collision are recorded at another moments. Numerical results reproduce the bifurcation transition process of tightly spaced SM from stationary phase to sliding phase during collision. To the best of our knowledge, it is the first in-deep study of the collision between dual-color optical SMs. Our results contribute to the understanding of the complex nonlinear interactions between solitons in dual-wavelength MLFLs.

## 2. Experimental setup

We set up an experimental device including a dual-wavelength dispersion-managed MLFL and a real-time DFT-based spectral measurement system to analyze the intra-cavity exotic dynamics, as shown in Fig. 1(a). Similar to our previous work [45], a dual-wavelength dispersion-managed MLFL mode-locked by NPR mechanism is used for dual-color soliton generation. A segment of 51-cm erbium-doped fiber (EDF, LIEKKI Er110-4/125), is backward-pumped by a 976 nm laser diode through a 980/1550 nm wavelength division multiplexer (WDM). The pigtail fiber of WDM is an 85-cm OFS980. The group-velocity dispersion (GVD) of the EDF and the OFS980 are +0.012 $ps^2$/m and +0.005 $ps^2$/m, respectively. The sandwich structure of two polarization controllers (PCs) and a polarization-dependent isolator (PD-ISO) is employed as an artificial saturable absorber. A 14-cm polarization maintaining fiber (PMF) provides the birefringence required for dual wavelength mode-locking, corresponding to the spectral filtering bandwidth of ~40 nm. In order to compensate the intra-cavity dispersion to near-zero, a segment of 2.1-m dispersion compensation fiber (DCF, Thorlabs DCF38) with the GVD of +0.048 $ps^2$/m is inserted into the ring cavity. The pigtails of optical components are all single-mode fiber (SMF) with the GVD of -0.022 $ps^2$/m, and the corresponding length is ~6.57 m. The overall fiber length of the ring cavity is 10.17 m corresponding to the net dispersion of −0.036 $ps^2$, which indicates that the ring laser operates at the near-zero anomalous dispersion. The pulses are output through a 10:90 output coupler (OC) from the ring cavity. Part of the output signals are sent directly to the optical spectrum analyzer (OSA, Yokogawa AQ6370B) for monitoring the current operation. The rest is used for shot-to-shot spectral evolution measurements.

In order to avoid the aliasing of the intra-cavity and extra-cavity collisions, the asynchronous pulse sequences with different center wavelengths emitted from the laser oscillator are separated by a bandpass filter. In Fig. 1(a), the pulses with short- and long-wavelengths are marked in blue and red, respectively. A segment of 4-m long SMF delay line is inserted into the branch of the red pulse. Therefore, an extra delay is introduced between the red and blue pulses. It is noteworthy that the time-delay operation is necessary to analyze the collision characteristic of dual-wavelength mode-locking. Then, the two spectra are combined again in a 3-dB fiber coupler, and pass through a 10-km DCF module with the dispersion parameter of -342 ps/nm in the C band together. The spectrum of the pulse is mapped by chromatic dispersion to a temporal waveform whose intensity envelope reflects the spectral information. The stretched pulses are detected by a 20-GS/s-sampling-rate real-time oscilloscope (OSC, LeCroy WaveRunner 640Zi, 4 GHz bandwidth) equipped with a fast photodetector (PD, ET-5000F, 12 GHz bandwidth), which renders the spectral resolution of 0.26 nm.

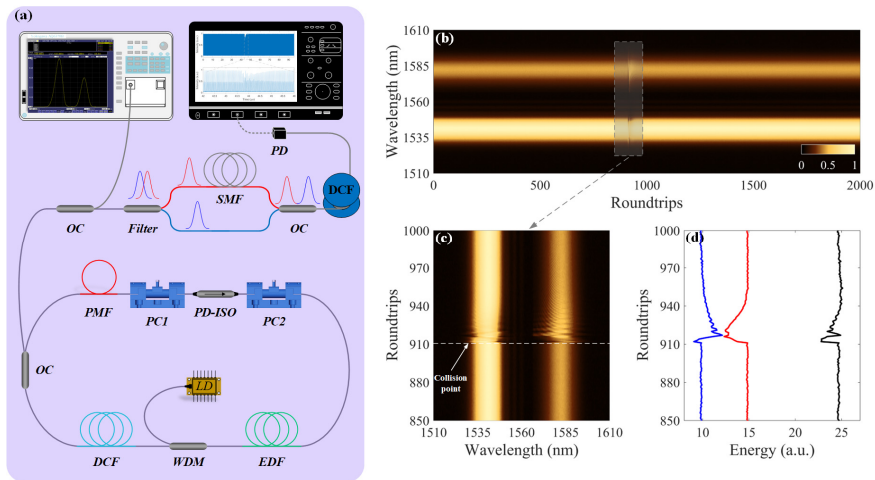

**Fig. 1.** (a) schematic of the dual-wavelength dispersion-managed MLFL and the real-time DFT-based measurement system; (b) real-time spectral evolution of the dual-wavelength soliton singlet mode-locking during 3000 roundtrips, and the collision happens at 911th roundtrip; (c) close-up of the collision process in (a); (d) energies of short- (marked in blue) and long-wavelength (marked in red) mode-locked pulses and their total energy (marked in black).

## 3. Experimental results and discussions

### 3.1. Collision between dual-color SSs

The inherent repetition frequency difference between dual-wavelength asynchronous pulses drives inevitable periodic collisions. Fig. 1(b) exhibits the real-time spectral evolution by DFT technique in 3000 roundtrips. The stable evolution is hardly affected by the collision, as can be seen from Fig. 1(d) that the energies before and after the collision are at the similar level. The collision and reconstruction process lasts about 60 roundtrips, accounting for only 0.13% of a collision period (corresponding to the repetition frequency difference of 443 Hz). Thus, it is not enough to influence the short-term stability of dual-wavelength mode-locking and its application in some precision measurement fields [46-48]. Surprisingly, contrary to the previous experimental evidences for dual-wavelength mode-locked collisions in the anomalous dispersion regime [42], the overlap of the pulse tails at the collision point causes their spectra to repulse each other instead of attracting, as shown in Fig. 1(c). The white dotted line marks the location of the collision point. The energy fluctuations during the collision process in Fig. 1(d) correspond to the spectrum first repulsing, then attracting and finally recovering from the asymmetric spectral structure. The total energy suddenly increases and then gradually dissipates after collision. In a dissipative system, attractive and repulsive forces between pulses are provided by cross-phase modulation (XPM) and cross-amplitude modulation (XAM), respectively [49,50]. In Ref. [42], XPM triggers dispersive wave shedding during the collision process. However, for sideband-free dispersion-managed solitons, XAM may act before XPM at the collision point, causing the spectral repulsion. Another appealing difference is the soliton interaction length during the collision. Dispersion management makes the interaction length more than one roundtrip, which will be illustrated later. Periodic collisions are the basis of our research on complex nonlinear dynamics in the dual-wavelength MLFL.

### 3.2. Collision between dual-color SMs

The energy boost in the oscillator facilitates the formation of multi-soliton patterns. In the near-zero dispersion regime, the interaction between solitons leading to the formation of bound states is originated from the direct overlap of stretched pulses [49]. The shot-to-shot real-time interferograms in Fig. 2(a) depict the smooth transitions of dual-color stationary SMs. Close-up of the internal motion of SMs during the collision process is shown in Fig. 2(b), in which their fast destabilization and reconstruction are clearly observed. For convenience, we define the short- and long-wavelength SMs as blue and red SMs, respectively. Taking the blue SM with sparse fringes as an example, the separation and intensity of the leading and tailing pulses undergo the extreme to slight oscillations, and recover to the initial stable point after a collision period. Generally, relative phase and temporal separation are considered as the degrees of freedom and intuitive representations of internal dynamics, dominating the intra-molecular temporal and spatial motions. The interaction planes, taking the temporal separation as the radius and the relative phase as the angle are used to reveal how dynamic attractors evolute during collision in the phase space [see Fig. 2(d) and 2(e)]. The blue SM fluctuates in the region near the fixed point before and after collision, while the red SM moves away from the previous fixed point after collision. The relative phases all fluctuate around the fixed point before and after collision, which are mainly originated from the quantum diffusion of the pulses caused by quantum-limited noises [51,52]. The collision breaks the relative phase-locking of bound solitons, appearing as scattered green dots in the interaction planes. The relative phase differences during collision of blue and red SMs are ~4π and ~3π, respectively, evident in Fig. 2(c).

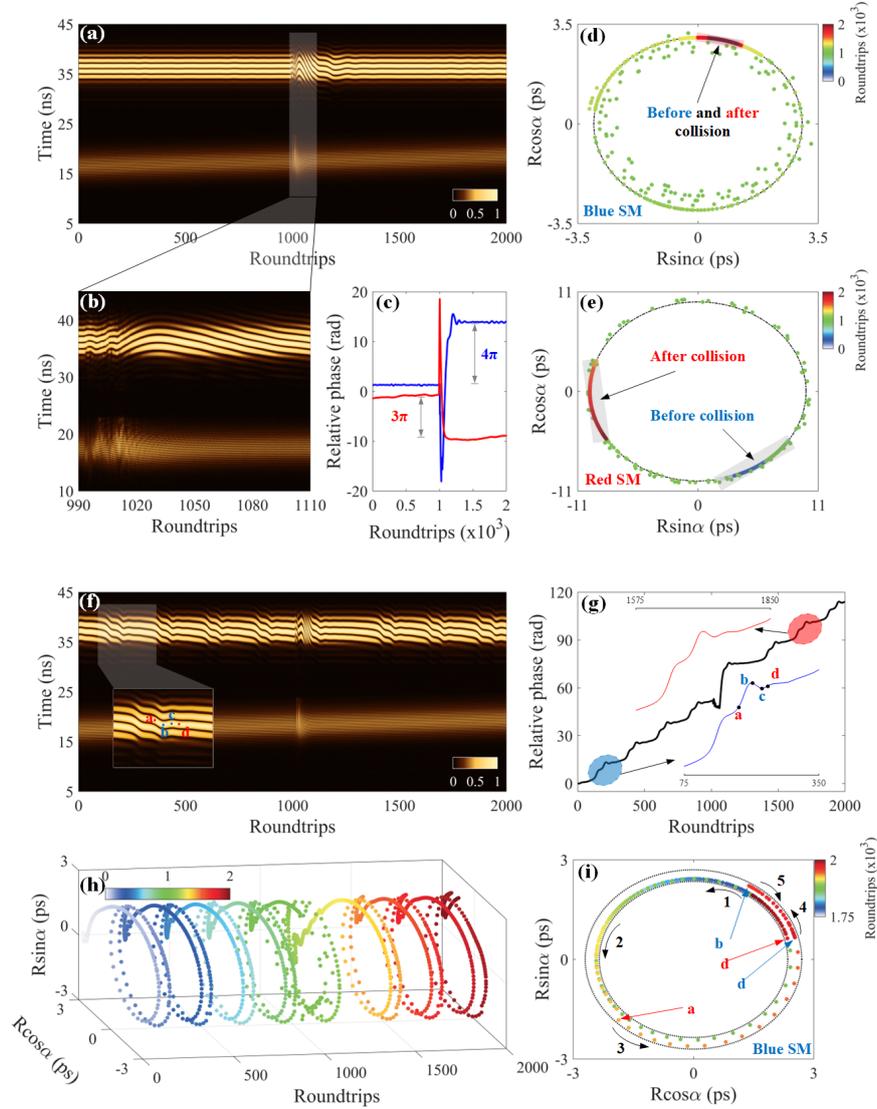

**Fig. 2.** Smooth transitions of dual-color temporal SMs during collision (for convenience, SMs with short- and long-wavelength are defined as blue and red SMs, respectively). (a)-(e) collision of dual-color stationary SMs; (a) experimental real-time interferograms; (b) close-up of the collision and reconstruction process in 120 roundtrips in (a); (c) relative phase fluctuations of blue and red SMs; (d), (e) the interaction plane of blue and red SMs, respectively; (f)-(h) collision between vibrating blue SM and stationary red SM; (f) experimental real-time interferograms, and the inset shows the close-up; (g) relative phase of blue SM, and the inset shows the close-up before (blue line) and after collision (red line); (h) 3D interaction space of blue SM; (i) the interaction plane of the blue SM between 1750 and 2000 roundtrips.

Beyond the phase-locked bound states, the collision referring to the vibrating SMs with evolving phase and pulse separation is also experimentally recorded, as shown in Fig. 2(f)-2(i). The close-up in Fig. 2(f) exhibits an entire sliding process including two separation jump points (a, d) and two phase turning points (b, c). To further resolve such vibrating SM with compound sliding phase, the relative phase evolution and interaction plane are plotted in Fig. 2(g) and 2(i). The blue and red lines in Fig. 2(g) are close-ups of relative phase evolutions before and after collision, respectively. The consistency of the undulation proves that the sliding period of the blue SM is not perturbed by collision. There is a 2-torus formed by the black dotted line in the interaction plane in which steps of the evolution are marked with numbers, as shown in Fig. 2(i). The relative phase jumps at a-point to the limit cycle with a larger separation (in contrast to the situation at d-point), and U-turns at b- and c-point on the same limit cycle. 3D interaction space of blue SM in Fig. 2(h) more intuitively exhibits the stepping phase evolution and flipping motions. Combining the relative phase evolution and the interaction plane [see Fig. 2(g) and 2(i)], the direction of the phase sliding is tracked from the evolution trajectories of successive interferograms [see the inset of Fig. 2(f)]: the spectral fringes shift towards longer wavelengths, corresponding to a counterclockwise rotation of the patterns in the interaction plane, indicating that the intensity of the tailing pulse is stronger than that of the leading pulse. This slight difference in intensity can be explained by their different phase velocities in dispersive medium induced by the intensity-dependent Kerr effect, and results in a different phase shift of the carrier envelope after each round trip [22]. This simple relative phase evolution changes dramatically at a-point. The temporal separations of the bound solitons from 2.43 ps to 2.61 ps causes a stepping speed change of the phase evolution, as shown by the slope between a- and b-points of blue line in Fig. 2(i). Subsequently, the intensity of the bound pulses alternately dominates, corresponding to changes in the shifting direction of the fringes. Finally, the separation is restored to 2.43 ps at d-point.

In the frame of nonlinear dynamics, the local instability of the fixed-point attractor leads to the appearance of a limit cycle attractor via a Hopf-type bifurcation. Considering the SMs in fiber lasers, the excitation of vibrating SM is generally achieved by the changes in pump power or polarization states. Instead, we observe the collision-induced Hopf-type bifurcation, in which energy exchange and complex nonlinear interaction are witnessed [37]. A clear and well-resolved reversible Hopf-type bifurcation transitions is shown in Fig. 3. The bifurcation transition during collision consists of four stages: fixed phase and separation (stage I), collision-induced extreme changes in phase and separation (stage II), separation-dominated phase sliding (stage III), and phase-dominated sliding (stage IV). The inverse process of bifurcation can also be divided in four stages, however phase-dominated sliding (stage I) and fixed phase and separation (stage IV) are opposite to that in bifurcation transition. The spectral evolutions and the corresponding shot-to-shot Fourier transform based first-order autocorrelation traces are divided according to the above four stages [see Fig. 3(a) and 3(b), 3(e) and 3(f)]. Since the change of pulse separation and phase of red SM during collision are consistent with that in Fig. 2, only the blue SM is discussed here. In the stage II of collision-induced extreme changes in phase and separation, the collision process of two-color SMs is more complicated due to the existence of intermolecular and intramolecular interactions. XPM induced by the overlap between molecules leads to the internal motions of SMs. The difference in the separation of bound pulses between SM with stationary phase and sliding phase originates from the phase space topology of different cavity parameters. In order to reflect the intramolecular relative phase evolutions clearly and intuitively before and after collision, discrete interaction points during collision are hidden [see Fig. 3(d) and 3(h)]. Different evolutions of stage II and stage III are manifested as a fast inward orbital transfer and a stepped outward orbital transfer in the interaction plane, respectively. The phase-dominated sliding of the inverse process of bifurcation is divided into two stages: double limit-cycle phase sliding which is similar to that Fig. 2(h) and 2(i) and phase sliding on a fixed limit cycle. In addition, the sign reversal of the slope in the stage of phase-dominated sliding illustrates a brief intensity reversal of bound pulses, as shown in Fig. 3(c) and 3(g).

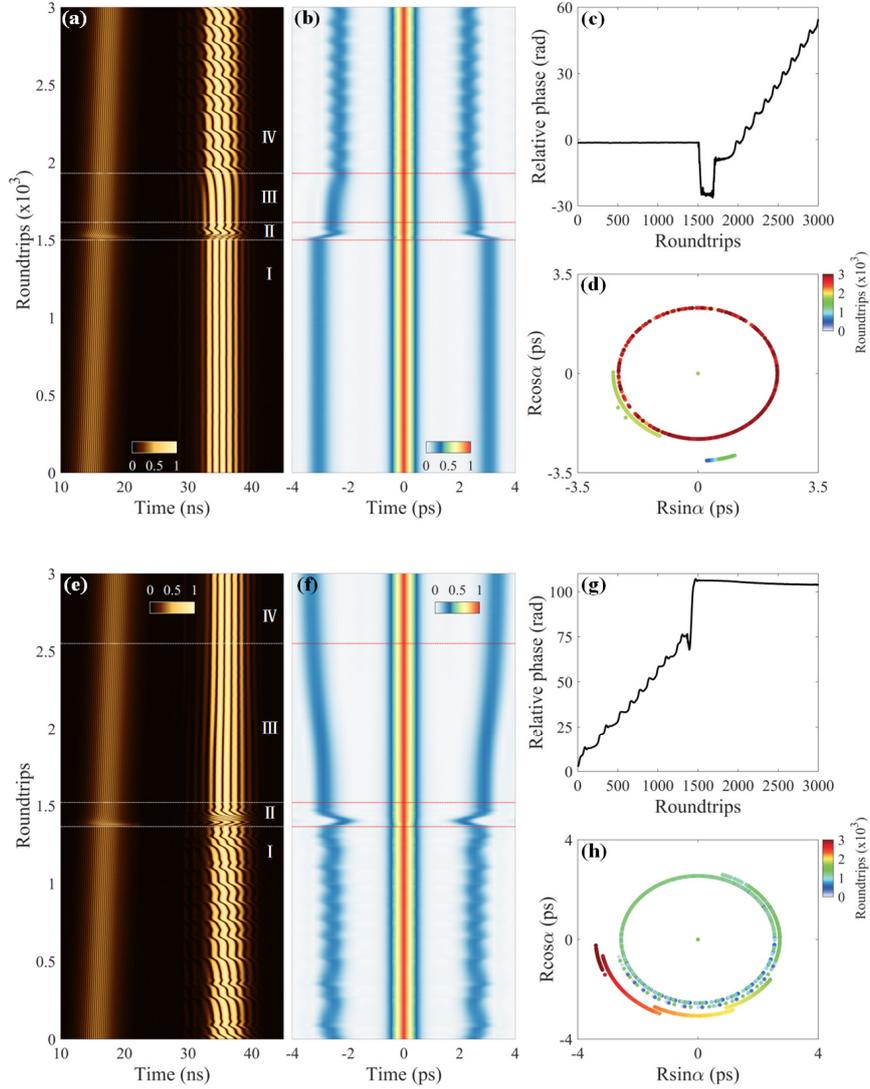

**Fig. 3.** Collision-induced bifurcation reversible transitions of dual-color temporal SMs (since the change of pulse separation and phase of red SM during collision are consistent with that in Fig. 2, the SM mentioned here only represents blue SM). (a)-(d) transition from stationary SM to vibrating SM during collision; (e)-(h) transition from vibrating SM to stationary SM during collision; (a), (e) experimental real-time interferograms; (b), (f) 2D contour plot of the shot-to-shot first-order autocorrelation traces; (c), (g) relative phase retrieved from autocorrelation traces; (d), (h) the interaction plane before and after collision.

Remarkably, collisions of the dual-color SMs above are obtained with the same cavity setups (the fixed PC orientations and pump power) [see Fig. 2 and Fig. 3]. We capture the smooth or reversible bifurcation transitions at different moments, revealing that the cavity parameters sometimes pass through the Hopf-type bifurcation point during collision. Specifically, the phase space of the cavity parameters randomly switches between the two or more different topologies through periodic collisions. In other words, collision can be considered as the intersection of different topologies. A specific observation captures only a projection of intracavity complex nonlinear dynamics at a certain moment. On the other hand, SMs with molecular-like properties are more sensitive to external interventions, such as changes of pump power and intracavity polarization state, so that the bonds between its internal constituents break through the critical stabilization and eventually cause the bifurcation [31]. However, the bifurcation evolution around the moment of external intervention is difficult to be captured. In contrast, it is not difficult to achieve for the intracavity inherent collision with obvious characteristics. Single-cavity dual-comb laser systems provide a fantastic platform for researching bifurcation dynamics.

### 3.3. Numerical simulation

In order to confirm the universality of collision-induced bifurcation behavior in single-cavity dual-comb laser systems, we numerically simulate the collision process of asynchronous optical pulses in a fiber laser. The simulation model is governed by the coupled Ginzburg-Landau equation (CGLE) [53] including the combination of fiber dispersion, transmission loss, fast saturable absorption, and gain saturation. To solve the CGLE, Euler method is used to compute the differential operator, and 4th order Runge-Kutta method is used to compute the integration in the propagation distance [54]. A two-peak gain spectral profile is integrated into the EDF to suppress gain competition and form dual-wavelength mode-locking. The propagation path of the pulses in the simulation completely follows the experimental structure shown in Fig. 1(a). Simulation results restore the bifurcation behavior of dual-color SMs at a higher resolution, as shown in Fig. 4. Because the Hopf-type bifurcation is random and uncontrollable, we do not reproduce the sliding phase dynamics of double limit cycles. Nevertheless, a proof-of-principle can still be obtained from the simulation results.

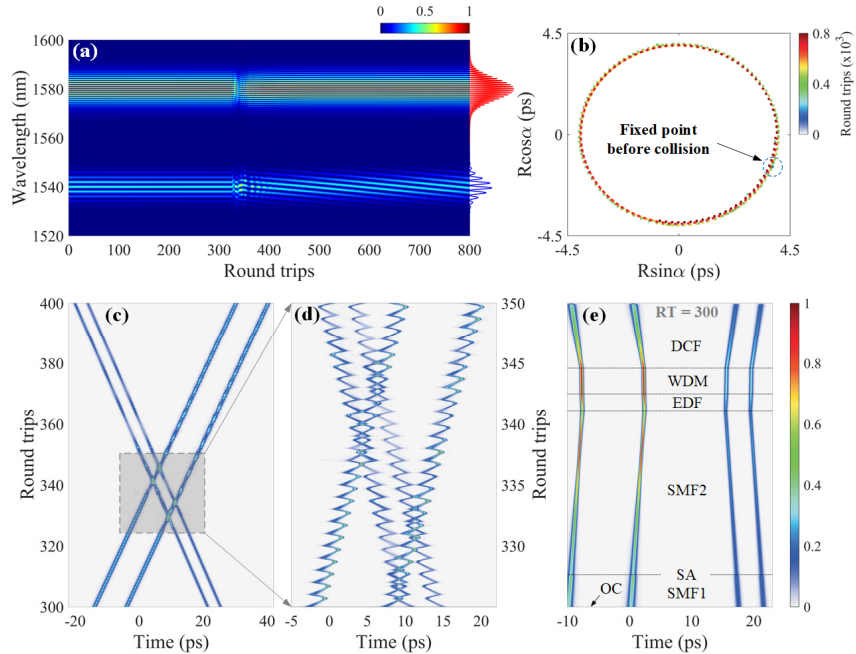

**Fig. 4.** Collision-induced bifurcation evolution of dual-color SMs in the numerical simulation. (a) the spectral evolution of dual-color SMs with different temporal separations during collision; (b) the interaction plane of blue SM; (c) the corresponding temporal evolution in 100 roundtrips; (d) close-up of the collision process in (c), evolution of asynchronous pulse pairs in successive 25 roundtrips; (e) the temporal evolution in a single roundtrip. SA: saturable absorber.

The phase of the blue SM slides linearly after the collision, while the red SM with large temporal separation still maintains a smooth transition as in the experiment, that is, no bifurcation behavior occurs [see Fig. 4(a)], revealing that the experimental results are not accidental. Energy exchange inside bound solitons becomes more important in the phase-dominated dynamics of tightly spaced pulses and less obvious for large pulse separation [26]. As a result, the stability of tightly spaced SMs is more likely to be destroyed during collision. The interaction plane in Fig. 4(b) depicts that the relative phase starts from a noise-free fixed point with locked relative phase and converting to a stepping evolution on a limit cycle after collision. Moreover, the opposing motion of dual-color pulses in Fig. 4(c) is attributed to the setting in numerical simulation model. The center wavelength of the model is 1560 nm, while the center wavelengths of dual-wavelength mode-locked pulses are 1540nm and 1580nm, respectively. In the simulation, we query the output for each location within the cavity to analyze the details of the collision process, evident in Fig. 4(d). Dispersion management enables pulses from different molecules to entangle together within several roundtrips, which is different from conventional solitons that collide completely within one roundtrip [42]. The temporal evolution in a single roundtrip illustrates that the 'boomerang' time shift of the pulses in Fig. 4(d) originates from the opposite-signed GVD of different fibers [see Fig. 4(e)].

## 4. Conclusion

To summarize, we report abundant exotic nonlinear phenomena in a dual-wavelength dispersion-managed ultrafast fiber laser. Different from traditional solitons with the short interaction length (no more than one roundtrip), dispersion management-dominated 'boomerang' time shift maintains the overlap between different color pulses for several roundtrips, making the collision process more complex and unpredictable. On this basis, experimental evidences of collisions between two-color SMs reveal the particle properties of optical SMs. The evolution direction of the internal degrees of freedom of the SM during collision process is random. We demonstrate the existence of collision-induced smooth or reversible transitions of Hopf-type bifurcations without artificially altering cavity parameters, indicating that the collision process sometimes passes through a bifurcation point. The numerical simulation results demonstrate the experimental observation that the tightly spaced SM is more prone to bifurcation. However, further researches are needed to verify whether the collision between dual-color SMs is similar to the anti-crossing collision between a soliton pair and a soliton singlet [6]. Despite the transient dynamics of collisions in a dual-color soliton fiber laser have been solved by TS-DFT technique, whether the two solitons or SMs exchange their eigenstates remains indistinguishable experimentally. Moreover, the ultracold molecule collision is an important research direction in condensed-matter physics [55,56]. Therefore, we expect our results to offer a novel insight into intermolecular interactions ubiquitous in nonlinear science.

**Funding.** Project of Guangdong Province, China (2018B090944001); National Natural Science Foundation of China (61827821, 61975144).

**Disclosures.** The authors declare no conflicts of interest.

**Data availability.** Data underlying the results presented in this paper are not publicly available at this time but may be obtained from the authors upon reasonable request.